\documentclass[amsmath,amssymb,nofootinbib,preprint]{revtex4}

\usepackage{latexsym,comment}
\usepackage{amssymb}
\usepackage{amsfonts}
\usepackage{amsmath,color}
\usepackage[dvips]{graphicx}
\usepackage{bm}

\def\mb#1{\mathbf{#1}}

\def\ber{\begin{eqnarray}}
\def\eer{\end{eqnarray}}
\def\beq{\begin{equation}}
\def\eeq{\end{equation}}

\def\ed{\end{document}}
\def\di#1#2{\frac{\mathrm{d} #1}{\mathrm{d}#2}}

\begin{document}

\title{Sagnac Effect, Ring Lasers and Terrestrial Tests of Gravity}

\author{Matteo Luca Ruggiero$^{\star,}$}
\email{matteo.ruggiero@polito.it}
\affiliation{$^\star$ Member of the GINGER collaboration \\ DISAT, Politecnico di Torino, Corso Duca degli Abruzzi 24,  Torino, Italy\\
 INFN, Sezione di Torino, Via Pietro Giuria 1, Torino, Italy}

\date{\today}

\begin{abstract}
Light can be used as a probe to explore the structure of space-time: this is usual in astrophysical and cosmological tests, however it has been recently suggested that this can be done also in terrestrial laboratories. Namely, the GINGER project aims at measuring post-Newtonian effects, such as the gravito-magnetic ones, in an Earth based laboratory, by means of a ring lasers array. Here, we first review the theoretical foundations of the Sagnac Effect, on which ring lasers are based, and then  we study the Sagnac Effect in a terrestrial laboratory, emphasizing the origin of the gravitational contributions that GINGER aims at measuring. Moreover, we show that accurate measurements allow to set constraints on theories of gravity different from General Relativity. Eventually, we describe the experimental setup of GINGER.
\end{abstract}

\maketitle

\section{Introduction}\label{sec:sdec}

2015 is the International Year of Light \cite{2015} and it celebrates the importance of light in science, technology and society development. As for physics, it is always useful to remember the central role of light in the genesis of the Theory of Relativity: in his \textit{Autobiographical Notes}, Einstein wrote that, at the age of 16, he imagined to chase after a beam of light and that this thought experiment had played a very important role in the development of Special Relativity. 2015 precedes by one year the centennial celebration of the publication of General Relativity (GR): it is interesting to emphasize that, except for the perihelion shift, the classical tests of GR exploit light as probe: think about the gravitational frequency shift, gravitational deflection, gravitational time delay. Indeed, the whole development of experimental gravitation (see e.g. \cite{Will:2014bqa}) testifies the connection between gravity and light. What can we say, almost 100 years later, about Einstein's theory of gravity? We know that, from one hand, GR has been verified with excellent agreement in the Solar System and in binary pulsar systems \cite{Will} but, on the other hand, its reliability is questioned by observations on large  scales, where the problems of dark matter and dark energy are still unsolved. To solve these issues and to try to reconcile gravity with quantum theory, several theories have been proposed that are alternative to GR or that extend Einstein's theory. Of course, these theories should agree with the known tests of GR: as a consequence, there are continuos improvements in tests of gravity, which are important also because there are features of GR that have not been fully explored up today, even though various attempts have been made: this is the case of the so-called gravito-magnetic (GM) effects  \cite{gem}. Actually, in GR a GM field is  generated by mass currents, in close analogy with classical electromagnetism: it is a known fact that 
the field equations of GR, in linear post-newtonian approximation, can be written in form of Maxwell equations for the gravito-electromagnetic (GEM) fields \cite{mashhoon01}, \cite{mashhoon03}.  {Indeed, the first derivation of the GEM field equations dates back to the pioneering works by H. Thirring\cite{thirring1,pfister1}.}  

Attempts of measuring GM effects have been performed in space only, up today  {(see \cite{Iorio:2010rk} for a review)}, by the LAGEOS satellites orbital analysis \cite{Ciufolini:2004rq,Ciufolini:2011},  {by using the MGS probe  \cite{iorio2006,iorio2010a}} and by the Gravity Probe B (GP-B) mission \cite{GP.B};  the LARES mission has been designed for this scope, and is now gathering data  \cite{LARES,LARESbis}.   {Some of such attempts are somewhat controversial for certain aspects, and raised a debate\cite{Iorio:2010rk,iorio2003a,renzetti2013b,renzetti2012,ciufolini2010,ciufolini2010x}.}

 {Actually,  the possibility  of testing GM effects in a terrestrial laboratory has been explored by various authors in the past (see e.g. \cite{braginsky1,braginsky2,cerdonio,iorio2003y,pascual}). In particular,} the use of light  is at the basis of the recently suggested possibility of testing GM and, more in general, post-Newtonian effects, for the first time, in a terrestrial laboratory  by means of an array of ring lasers \cite{GINGER11}: this is  the Gyroscopes In GEneral Relativity (GINGER) project \cite{GINGER14,gingerweb}. A ring laser gyro, or simply ring laser, is a rotation sensor  based on the Sagnac Effect: the latter consists in  the shift of the interference pattern arising when an interferometer is set into rotation, with respect to what is observed when the device is at rest\cite{sagnac05,sagnac13}.  More in general, the Sagnac Effect is  an observable consequence of   {the non-isotropy of the coordinate velocity of light,  related to the synchronization gap along a closed path in non-time-orthogonal frames (see e.g. \cite{found}, \cite{RRinRRF}, \cite{rizzi03a}, \cite{rizzi03b}, \cite{LL}, and references therein)}, both in flat and in curved space-time. The Sagnac Effect has become important since the development of lasers, which allowed a remarkable advance of light interferometry\cite{post67}. Today, there are several technological applications based on the Sagnac Effect, such as fiber optic gyroscopes, used in inertial navigation, and  ring laser gyroscopes, used in geophysics\cite{chow85, vali76, stedman}.  

As suggested in \cite{StedLT,GINGER11,ginger12,GINGER14} ring lasers may be used for measuring the gravito-magnetic effects of the Earth: in fact, these devices measure with great accuracy the rotation rate of the terrestrial laboratory where they are located with respect to an inertial frame, e.g. with respect to fixed stars and, in doing so, the gravitational drag of inertial frames comes into play (see for instance  \cite{MTW,ciufoliniwheeler}). 

In this paper, we aim at reviewing the conceptual bases of the use of ring lasers to probe the space-time structure. We start, in Section \ref{sec:sagnac}, by studying the Sagnac Effect, in arbitrary stationary space-time \cite{sagnac14}. In particular, we  show that the Sagnac Effect does not depend on the physical nature of the propagating beams, provided that suitable kinematic conditions are fulfilled; thus, it can be interpreted as a  property of the space-time structure itself \cite{sagnac14bis}. Moreover, we  discuss to what extent  the classical Sagnac Effect formula is independent from the position of the center of rotation and from the shape of the enclosed area for an Earth-bound interferometer and, also, its relation with the Aharonov-Bohm effect.
 Since we are interested in the Sagnac Effect in a terrestrial laboratory, in Section \ref{sec:STLab} we focus on the definition of the space-time metric in the   {frame where the interferometer is at rest} and then, in Section \ref{sec:SELab} we study the outcome of a Sagnac experiment, performed on the Earth, starting from a somewhat generic expression of the terrestrial gravitational field, in terms of a suitable Post-Newtonian Parameterized  (PPN) space-time metric. We give the explicit results for the case of GR, and evaluate the order of magnitude of the relevant contributions. In Section \ref{sec:altSagnac}, we briefly discuss the impact that alternative or extended theories of gravity  have in such experiments, thus suggesting the possibility that accurate measurements could help to constrain  these theories. In Section \ref{sec:disconc}, we introduce the GINGER project, discussing the key features of the proposed experiment; conclusions are eventually drawn in Section \ref{sec:conc}.

\section{The Sagnac Effect}\label{sec:sagnac}

Ring lasers are based on the Sagnac Effect. In this Section we review in details the principle of the Sagnac
Effect from a space-time perspective, in order to show its \textit{universality}: we mean that the effect is the same, independently of the nature of the interfering beams. A thorough approach for light interferometers can be found in \cite{kajari}, while for a description of what happens in general, both for light and matter beams see e.g. \cite{scorgie},\cite{sagnac14bis}.

At the beginning of last century, Sagnac first predicted and then verified that there is a shift of the interference pattern when an interferometer is set into rotation, with respect to what is observed when the device is at rest \cite{sagnac05,sagnac13}. If we denote by
 $\bm{\Omega }$  the (constant) rotation rate of the interferometer with respect to an inertial  frame, by $\mathbf{S}$  the vector associated to the area enclosed by the light path and by $\lambda $  the wavelength of light, the expected and measured fringe shift is $\displaystyle
\Delta z=4\frac{\mathbf{\bm{\Omega} \cdot S}}{\lambda c}$. Then, it is possible to obtain the proper time difference, which turns out to be
\begin{equation}
\Delta t=\frac{\lambda }{c}\Delta z=4\frac{\mathbf{\bm{\Omega} \cdot
S}}{c^{2}} \label{eq:sagnac2}
\end{equation}

From a historical perspective, it is interesting to point out that Sagnac himself interpreted the results of his experiments in support of the ether theory against the Special Theory of Relativity (SRT): in fact,  he wrote that   \textit{``[...] the observed interference effect turns out to be the optical vortex effect due to the motion of the system with respect to the ether [...]''}\cite{sagnac13}. Subsequently, some authors agreed with this reading of his seminal experiments  (see e.g. the review \cite{RRinRRF} and the other papers  on the subject in the monograph \cite{RRF}), thus suggesting that there could be a problem when SRT is applied to rotating reference frames. Actually, this is not the case: the Sagnac Effect can be completely explained in relativistic framework both in flat and curved space time (see e.g. \cite{found}, \cite{RRinRRF}, \cite{rizzi03a}, \cite{rizzi03b}, \cite{LL}, \cite{sagnac14}, \cite{sagnac14bis}), as we are going to show.

To fix the ideas, let us consider the following experimental setup: an interferometer is at rest in a reference frame (the \textit{interferometer frame}), and it simultaneously emits two beams:  they propagate in opposite directions in the same path and reach the emission point at different times: we  call \textit{Sagnac time delay} the proper time difference between the two times of arrival, measured in the interferometer frame. Our approach is quite general, since it is purely geometric, and can be applied both in flat and curved space-time: for instance, the interferometer frame could be a turntable in the laboratory, or a more general frame, such as a terrestrial laboratory, where the rotation effects have both kinematic and gravitational origin. 

In the interferometer frame we can choose a set of adapted   coordinates  $\{x^{\mu}\}=\{x^{0},x^{i}\}=\{ct,\mb x\}$ and we can write the squared line-element in the form \footnote{We use the following notation: Greek (running from 0 to 3) and Latin (running from 1 to 3) indices  {denote} space-time and spatial components, respectively;
letters in boldface like $\mb x$ indicate spatial vectors.} 
\beq
ds^{2}=g_{00}c^{2}dt^{2}+2g_{0i}cdtdx^{i}+g_{ij}dx^{i}dx^{j} \label{eq:metricastazionaria}
\eeq

Since we suppose that the space-time is stationary, the metric  does not depend on time: accordingly,  $g_{\mu\nu}=g_{\mu\nu}(\mb x)$; we choose the signature $(-,+,+,+)$, so that $g_{00}<0$; the above metric is not \textit{time-orthogonal}, because $g_{0i} \neq 0$.

We study time-like (for matter beams) and light-like (for light beams) particles propagating in the space-time given by Eq. (\ref{eq:metricastazionaria}), to calculate the time intervals needed for a complete round trip by the two beams; then, we obtain the difference between them. We use this notation:  in the 3-dimensional space  of the metric (\ref{eq:metricastazionaria}), the length element is
\beq
d\ell^{2}=g_{ij}dx^{i}dx^{j}  \label{eq:dl2}
\eeq
Then, the  particles have unit tangent vectors
\beq
\ell^{i}=\di{x^{i}}{\ell}, \label{eq:defell1}
\eeq
and we may write the components of the coordinate speed in the form
\beq
u^{i} = \di{x^{i}}{t} = u \ell^{i} \label{eq:defiu1}
\eeq
with
\beq
u^{2}={u^{i}u_{i}=g_{ij}u^{i}u^{j}}=\di{\ell^{2}}{t^{2}} \label{eq:defiu2}
\eeq

Let $d\tau$ be the proper time interval, measured by the particles along the path; of course, $d\tau$ can be zero for light beams. On substituting this expression in the line-element (\ref{eq:metricastazionaria}), taking into account (\ref{eq:defell1}), we obtain

\beq
-c^{2} d\tau^{2}=ds^{2}= g_{00}c^{2}dt^{2}+2g_{0i}\ell^{i}cdtd\ell+d\ell^{2}
\label{eq:metricaell1}
\eeq

Now, we have to impose a condition to say that the particles propagating in the two opposite directions are identical but differ only for the direction of propagation: intuitively, we have to say that the two particles beams have the same speed. However, the coordinate speed $u$ has not a direct physical meaning. If we want to give an operational meaning to the speed of a particle, we may proceed as follows. Let us consider the coordinate point of the interferometer frame, occupied by the particle at a given time; we introduce an inertial frame, relative to which this point  {is at rest:} this is the so-called \textit{Locally Co-Moving Inertial Frame} (LCIF).  In this frame, the proper element of distance  $d\sigma$ and time $dT$ can be defined in terms of the metric elements and coordinates intervals in the interferometer frame by (see \cite{LL, RRinRRF})
\beq
d\sigma^{}=\sqrt{\left(g_{ij}-\frac{g_{i0}g_{j0}}{g_{00}} \right)dx^{i}dx^{j}},\quad dT^{}=-\frac 1 c \frac{g_{\mu 0}}{\sqrt{-g_{00}}}dx^{\mu}
\eeq
Indeed, on using these expressions, the line-element (\ref{eq:metricastazionaria}) is locally Minskoskian in the form:
\beq
ds^{2}=d\sigma^{2}-c^{2}dT^{2} \label{eq:metricadsigmadT}
\eeq
In the LCIF an observer  attributes to a particle a speed of magnitude  $v=\di{\sigma}{T}$, i.e. the ratio between the proper element of distance $d\sigma$ traveled in a proper time interval $dT$, and $dT$. In doing so, we have been able to introduce the particle speed $v$, which has a well defined operational meaning. On substituting in (\ref{eq:metricadsigmadT}), we get
\beq
ds^{2}= \left (1-\frac{c^{2}}{v^{2}} \right) d\sigma^{2} =  \left (1-\frac{c^{2}}{v^{2}} \right) \left(g_{ij}-\frac{g_{i0}g_{j0}}{g_{00}} \right)dx^{i}dx^{j}
\eeq
Eventually, on taking into account Eqs. (\ref{eq:defell1}),(\ref{eq:defiu1}),(\ref{eq:defiu2}) and (\ref{eq:metricaell1}), we obtain
\beq
 \left (1-\frac{c^{2}}{v^{2}} \right) \left[1-\frac{\left(g_{0i}\ell^{i} \right)^{2}}{g_{00}} \right]d\ell^{2}= g_{00}(\mb x)c^{2}dt^{2}+2g_{0i}(\mb x)\ell^{i}cdtd\ell+d\ell^{2}
 \label{eq:dt1}
 \eeq
Eq. (\ref{eq:dt1}) can be solved for the coordinate  time interval $dt$; to this end, we introduce $\displaystyle \beta \doteq v/c$ and $\lambda \doteq g_{0i}\ell^{i}$. Notice that for light-like particles, on setting $ds^{2}=0$, we get  $\beta=1$, in agreement with the second postulate of Special Relativity,  and the left hand side of Equation (\ref{eq:dt1}) is equal to zero.  Eq. (\ref{eq:dt1}) now reads
\beq
 \left (1-\frac{1}{\beta^{2}} \right) \left[1-\frac{\lambda^{2}}{g_{00}} \right]d\ell^{2}= g_{00}(\mb x)c^{2}dt^{2}+2\lambda(\mb x)cdtd\ell+d\ell^{2}
 \label{eq:dt11}
\eeq
and we obtain the two solutions
\beq
dt_{\pm}= \frac{1}{|g_{00}|c} \left(\lambda d\ell \pm \frac{1}{\beta}\sqrt{\lambda^{2}+g_{00}}d\ell \right) \label{eq:dtsol1}
\eeq

Eq. (\ref{eq:dtsol1}) can be integrated along the propagation path  to obtain the coordinate time interval. We are interested in the future oriented branch of the light cone: hence we obtain two solutions,  corresponding to the propagation times along opposite directions in the path.  The speed of the particles, that could be different for particles having different physical nature, enters Eq. (\ref{eq:dtsol1}) through the $\beta$ coefficient only. The coordinate time intervals for the propagation in two opposite directions along the \textit{same}  path $\ell$ can be written as
\beq
t_{+} = \oint_{\ell} dt_{+}, \quad t_{-} =-\oint_{\ell} dt_{-} \label{eq:dtpdtm}
\eeq
So, the difference between the co-rotating ($t_{+}$) and counter-rotating ($t_{-}$) propagation times turns out to be
\beq
\Delta t = t_{+}-t_{-}= \oint_{\ell} \left(dt_{+}+dt_{-} \right) \label{eq:deltapm11}
\eeq
Now, we are able to impose a condition on the speeds of the particles: in particular, the expression of the time difference $\Delta t$ simplifies if we assume that \textit{ the speed $v$ (or equivalently $\beta$) is a function only of the position along the path}; the case $v=\mathrm{constant}$ along the path is a particular sub-case. This amounts to saying  that, in any LCIF along the path, the co-rotating and the counter-rotating beam have the same speed $v$ in opposite directions. When  this condition is fulfilled,  the coefficient in the second term in (\ref{eq:dtsol1}) is the same for both the co-rotating and the counter-rotating beam,  and we obtain
\beq
\Delta t = t_{+}-t_{-} = \frac 2 c \oint_{\ell} \frac{\lambda d\ell}{|g_{00}|}=-\frac 2 c \oint_{\ell} \frac{g_{0i}dx^{i}}{g_{00}} \label{eq:formulafond}
\eeq

In summary,  particles take different times for propagating along the path, depending on their speed,  but \textit{ the
difference between these times is always given by eq. (\ref{eq:formulafond}), in any stationary space-time, and for arbitrary paths,} both for matter and light particles, independently of their physical nature. Actually, there are several experiments that corroborate this result (see e.g. \cite{zimmermann65,atwood84,riehle91,hasselbach93,werner79}). From a theoretical viewpoint, it is related to the issue of the round-trip synchronization in frames that are not time-orthogonal: in fact,  the above condition on the particles speed holds in a LCIF, where clocks are Einstein-synchronized \cite{RRinRRF,RRS}. 

Once that the coordinate time difference is known, it is possibile to calculated the proper time difference in the interferometer frame.
If the interferometer is located at $P$,   the proper time difference that expresses  the \textit{Sagnac time delay} is 
\beq
\Delta \tau=-\frac 2 c \sqrt{g_{00}(\mb x_{P})}  \oint_{\ell} \frac{g_{0i}(\mb x)}{ g_{00}(\mb x)}  dx^{i} \label{eq:deltataulocal1}
\eeq
which is referred to as the \textit{Sagnac Effect.}

The Sagnac Effect (\ref{eq:sagnac2}) is expressed in terms  of area enclosed by the path of the beams; this leads to the analogy with the Aharonov-Bohm effect (see e.g. \cite{rizzi03a}). In order to write the time delay (\ref{eq:deltataulocal1}) in terms  of area enclosed by the path of the beams, we proceed as follows.  We define the vector field $\displaystyle \mb h (\mb x) \doteq g_{0i}(\mb x)$, and the scalar field\footnote{In particular $\mb h(\mb x)$ and $\varphi(\mb x)$ are a vector and a scalar with respect to the coordinate transformation $x'^{i}=x'^{i}(x^{i})$,  internal to the reference frame. } $\displaystyle \varphi(\mb x) \doteq \frac{1}{g_{00}(\mb x)}$, by which we may write the Sagnac time delay in the form
\beq
\Delta \tau=-\frac 2 c \sqrt{\frac{1}{\varphi(\mb x_{P})}}  \oint_{\ell} \varphi \mb h  \cdot d \mb x \label{eq:deltataulocal2}
\eeq
When we study the gravitational field of  rotating objects, in the so called gravito-electromagnetic formalism (see e.g. \cite{gem,mashhoon01}),  $\mb h(\mb x)$ is usually referred to as the \textit{gravito-magnetic} potential, which enables to formally introduce the \textit{gravito-magnetic field} $\mb b(\mb x) \doteq \bm \nabla \wedge \mb h(\mb x)$. 

By using the Stokes theorem, we may write the integral in (\ref{eq:deltataulocal2}) in the form
\beq
\oint_{\ell} \varphi \mb h  \cdot d \mb l = \int_{S} \left[\bm \nabla \wedge (\varphi \mb h) \right] \cdot d \mb S \label{eq:stokes1}
\eeq
where   $\mathbf S$ is the area vector of the surface enclosed by path of the beams. After some vector algebra,  we  eventually obtain
\beq
\Delta \tau=-\frac 2 c \sqrt{\frac{1}{\varphi (\mb x_{P})}} \int _{S} \left[\bm  \nabla \varphi(\mb x) \wedge \mb h  (\mb x)\right] \cdot d \mb S -\frac 2 c \sqrt{\frac{1}{\varphi (\mb x_{P})}}  \int _{S} \left[\varphi (\mb x)  \mb b (\mb x) \right] \cdot d \mb S \label{eq:sagnac2area}
\eeq
Eq. (\ref{eq:sagnac2area}) is the general form of the Sagnac Effect, for both matter and light beams, in terms of surface integrals. It is now useful to comment on this result, in connection with the purported analogy with the Aharonov-Bohm effect, according to which the Sagnac Effect can be described in terms of the flux of the  field  {$\mathbf b(\mb x)$} across the interferometer area. We see that this analogy is true   if $\varphi(\mb x)$ is constant over $S$ or its change is negligibly small. Moreover, it is important to emphasize  that,    in the case of the Aharonov-Bohm effect, the {magnetic field} is null along the trajectories of the particles, while in the Sagnac Effect the field  {$\mathbf b(\mb x)$ is not null. 

\section{The Space-Time in the  {Interferometer} Frame} \label{sec:STLab}

In this Section, we are going to define the space-time metric in the frame  where the interferometer is at rest,  {that is, our interferometer frame; notice that, for a terrestrial experiment like GINGER (see Section \ref{sec:disconc}), this frame corresponds to the laboratory frame.} To this end, we shall use the construction of the ``proper reference frame'' as described in Ref. \cite{ciufoliniwheeler,MTW}. We consider an observer, at rest in the  {interferometer} frame,  arbitrarily moving in a 
background space-time; we write the corresponding local metric in a neighborhood of its world-line\footnote{From now on, for the sake of clarity, we use units such that $c=1$; physical units will be restored in the following Section.}
(see e.g. \cite{MTW})
\begin{eqnarray}
g_{(0)(0)}&=&1+2 \bm{\mathcal{A}} \cdot \bm x+O(x^{2}), \label{eq:G00a} \\
g_{(0)(i)}&=&\Omega_{(i)(k)}x^{(k)}+O(x^{2}),  \label{eq:G0i} \\
g_{(i)(j)}&=&\eta_{(i)(j)}+O(x^{2}). \label{eq:Gij}
\end{eqnarray}
The above expressions of the space-time metric hold  near the world-line of the observer only,
where quadratic displacements terms are negligible. We suppose that the observer is provided with an orthonormal tetrad (parentheses refer to tetrad indices)
$ e_{(\alpha)}$, whose four-vector $ e_{(0)}$  coincides with his four-velocity $\mathcal{U}$, while
the four-vectors $ e_{(i)}$ define the basis of the spatial vectors in the tangent space along its
world-line. By construction we have $e_{(\alpha)} e_{(\beta)}=\eta_{(\alpha)(\beta)}$, where $\eta_{(\alpha)(\beta)}$ is the Minkowski tensor. The metric
components (\ref{eq:G00a})-(\ref{eq:Gij}) are expressed in  coordinates that are associated to the
given tetrad, namely the space coordinates $x^{(i)}$ and the observer's proper time $x^{(0)}$.
In the above equations, $\bm{\mathcal{A}}$ is the spatial projection
of the observer's four-acceleration, while  the tensor $\Omega_{(i)(k)}$ is related to the parallel
transport of the basis four-vectors along the observer's world-line: $ \nabla_{\mathcal{U}}
e_{(\alpha)}=- e_{(\beta)}\Omega^{(\beta)}_{\ (\alpha)}$. In particular, if $\Omega_{(i)(j)}$ were zero,
the tetrad would be Fermi-Walker transported. Let us remark that the metric (\ref{eq:G00a})-(\ref{eq:Gij}) is
Minkowskian along the observer's world-line ($x^{(i)}=0$); it is everywhere flat iff $\bm{\mathcal{A}}=0$,
i.e. the observer is in geodesic motion and the tetrad is non rotating (i.e. it does not rotate with
respect to an inertial-guidance gyroscope).  In the latter case, the first corrections to the flat
space-time metric are $O(x^{2})$ and are proportional to the Riemann tensor  \cite{MTW}.  {In what follows, we assume that the space-time metric $g_{\mu\nu}$ of the interferometer frame, which we used in Section \ref{sec:sagnac} to obtain the explicit expression (\ref{eq:sagnac2area}) of the Sagnac effect, is the given by the local metric  (\ref{eq:G00a})-(\ref{eq:Gij}).}

In order to explicitly write the local metric, which through its  gravito-magnetic ($g_{0i}$)  and gravito-electric ($g_{00}$) components enables us to evaluate the Sagnac Effect,  we must choose a suitable tetrad by taking into account the motion of the Earth-bound laboratory in the background space-time metric. We consider the following PPN background metric which describes the gravitational field of the
rotating Earth (see e.g. \cite{Will}):
\begin{eqnarray}
&ds^{2}&= {G_{\mu\nu}\, dX^{\mu} dX^{\nu}}=(1-2U(R))dT^{2}-\left(1+2\gamma U(R)\right)\delta_{ij}dX^{i}dX^{j}+ \nonumber \\
&2&\!\!\!\!\!\!\left[\frac{\left(1+\gamma+\alpha_{1}/4\right)}{R^{3}}{\left(\bm J_{\oplus}
\wedge \bm R\right)_{i}}-\alpha_{1}U(R) W_{i}\right]dX^{i}dT , \nonumber \\
&\ &
\label{eq:metricappn}
\end{eqnarray}
where $-U(R)$ is the Newtonian potential, $\bm J_{\oplus}$ is the angular momentum of the Earth,
$W_{i}$ is the velocity of the reference frame in which the Earth is at rest with respect to mean
rest-frame  of the Universe; $\gamma$ and $\alpha_{1}$ are post-Newtonian parameters that measure,
respectively, the effect of spatial curvature and the effect of preferred frames. 

The background metric (\ref{eq:metricappn}) is referred to an Earth Fixed Inertial (ECI) frame,
where Cartesian geocentric coordinates are used, such that $\bm R$ is the position vector
and $R \doteq \sqrt{\sum_{i} X^{2}_{i}}=\sqrt{X^{2}+Y^{2}+Z^{2}}$. Then, we choose a laboratory
tetrad which is related to the background coordinate basis of (\ref{eq:metricappn}) by a pure Lorentz
boost, together with a re-normalization of the basis vectors: in other words the local laboratory  axes have the same orientations as those in the background ECI frame, and they could be physically realized by three orthonormal telescopes, always pointing toward the same distant stars.  In this case,  one can  show\cite{MTW,ciufoliniwheeler,tourrenc,ashby90}  that  in the local metric 
$\Omega_{(i)(k)}x^{(k)}=-\left(\bm{\Omega}^\prime \wedge \bm x \right)_{(i)}$, where
the total relativistic contribution $\bm  \Omega'$ is the sum of four terms, with the  dimensions of angular rotation rates
\beq
\bm \Omega' = \bm \Omega_{G}+ \bm \Omega_{B}+  \bm \Omega_{W} +\bm\Omega_{T} \label{eq:omegaprime}
\eeq
defined by
\begin{eqnarray}\bm \Omega_{G}&=&
-\left( 1+\gamma \right) \bm \nabla U(R) \wedge \bm V, \label{eq:OmegaDS} \\
\bm \Omega_{B}&=&-\frac{1+\gamma +\alpha_{1}/4}{2} \left(\frac{\bm J_{\oplus}}{R^{3}}-\frac{3 \bm
 J_{\oplus} \cdot \bm R}{R^{5}}\bm R \right),\label{eq:OmegaLT} \\
\bm \Omega_{W} & = & \alpha_{1}\frac{{1}}{4} \bm \nabla U(R) \wedge \bm W, \label{eq:Omegaw} \\
\bm \Omega_{T}&=&-\frac{1}{2} \bm V \wedge  \frac{d \bm V}{dT}. \label{eq:OmegaTh}
\end{eqnarray}
Indeed, the vector $\bm  \Omega'$ represents the precession rate that  an inertial-guidance gyroscope, co-moving with the laboratory, would have with respect to the  \textit{ideal} laboratory  spatial axes (see e.g. \cite{MTW,ciufoliniwheeler}) which are always oriented as those of the ECI frame. Differently speaking, we may say that the local spatial basis vectors  are not Fermi-Walker transported along the laboratory world-line.
 In detail, the total precession rate is made of four contributions:
 i) the geodetic or de Sitter precession $\bm \Omega_{G}$ is due to the motion of the laboratory
in the curved space-time around the Earth; ii) the Lense-Thirring or gravito-magnetic precession $\bm\Omega_{B}$ is due to
the angular momentum of the Earth; iii)  $\bm \Omega_{W}$ is due to the preferred frames effect;
and iv) the Thomas precession $\bm \Omega_{T}$ is  related to the angular defect due to the Lorentz boost. It is worth noticing that  for a laboratory  bounded to the Earth
\beq
\bm{\mathcal A}\simeq\frac{d\bm V}{dT}-\bm{\nabla}U(R), \label{eq:Alab}
\eeq
and the  acceleration $\bm{\mathcal A}$ can not be eliminated.
We emphasize that all  terms in (\ref{eq:OmegaDS})-(\ref{eq:OmegaTh}) must be evaluated along the laboratory world-line (hence, they are constant in the local frame), whose position and velocity in the background frame are $\bm R$ and $\bm V$, respectively.
However, if we consider an \textit{actual} laboratory fixed on the Earth surface, the spatial axes of the
corresponding tetrad rotate with respect to the coordinate basis of the metric (\ref{eq:metricappn}),
and we must take into account in the gravito-magnetic term (\ref{eq:G0i})
the contribution of the additional rotation vector $\bm \Omega_{\oplus} $, which corresponds to the Earth
rotation rate, as measured in the local frame\footnote{For an Earth-bounded laboratory, it is
$\bm \Omega_{\oplus} \simeq \left[1+U(R)+\frac{1}{2}\Omega^{2}_{0} R_{}^{2}  \sin^{2} \vartheta \right]\bm \Omega_{0}$ 
where $R_{}$ is the terrestrial radius,  $\vartheta$ is the colatitude angle of the laboratory and $\bm \Omega_{0}$ is the terrestrial rotation rate, as measured in an asymptotically flat inertial frame.}.

As a consequence,  it is possible to show that, up to linear displacements from the world-line, the off-diagonal term in the metric can be written as
\beq
g_{(0)(i)}= \left(\bm{\Omega } \wedge \bm x \right)_{(i)},
\label{eq:Omegaik2}
\eeq
where $\bm \Omega=-\bm \Omega_{\oplus}-\bm \Omega'$.

\section{The Sagnac Effect in the  {Interferometer} Frame}\label{sec:SELab}

We are able to evaluate the proper-time difference

\beq
\Delta \tau=-2  \sqrt{\frac{1}{\varphi (\mb x_{P})}} \int _{S} \left[\bm  \nabla \varphi(\mb x) \wedge \mb h  (\mb x)\right] \cdot d \mb S - 2  \sqrt{\frac{1}{\varphi (\mb x_{P})}}  \int _{S} \left[\varphi (\mb x)  \mb b (\mb x) \right] \cdot d \mb S \label{eq:sagnac2areabis}
\eeq
taking into account the general expressions  (\ref{eq:G00a})-(\ref{eq:Gij}) of the space-time metric in the  {interferometer} frame.  It is possible to apply our general relation (\ref{eq:sagnac2area}). In this case, it is
\beq
\varphi(x^{i})=\frac{1}{1+2\bm{\mathcal A} \cdot \mb x}, \quad \mb h(x^{i})= \left(\bm{\Omega } \wedge \mb x \right) \label{eq:defphhlab}
\eeq
In particular, we see that   $\mb b= 2 \bm \Omega$.  On substituting in (\ref{eq:sagnac2areabis}) we obtain
\beq
\Delta \tau=-2 \sqrt{1+2\bm{\mathcal A} \cdot \mb x_{P}} \int _{S} \left[\frac{-2 \bm{\mathcal A} \wedge  \left(\bm{\Omega } \wedge \mb x \right) }{\left(1+2\bm{\mathcal A} \cdot \mb x \right)^{2}} \right] \cdot d \mb S -2 \sqrt{1+2\bm{\mathcal A} \cdot \mb x_{P}}  \int _{S} \left[\frac{\bm \nabla \wedge \left(\bm{\Omega } \wedge \mb x \right) }{1+2\bm{\mathcal A} \cdot \mb x} \right] \cdot d \mb S \label{eq:sagnaclocal11}
\eeq
and then, by performing the vector operations
\beq
\Delta \tau=4 \sqrt{1+2\bm{\mathcal A} \cdot \mb x_{P}} \int _{S} \left[\frac{\bm \Omega \left( \bm{\mathcal A} \cdot \mb x \right)- \mb x \left(\bm{\mathcal A} \cdot \bm \Omega \right)}{\left(1+2\bm{\mathcal A} \cdot \mb x \right)^{2}} \right] \cdot d \mb S -4 \sqrt{1+2\bm{\mathcal A} \cdot \mb x_{P}}  \int _{S} \left[\frac{\bm \Omega }{1+2\bm{\mathcal A} \cdot \mb x} \right] \cdot d \mb S \label{eq:sagnaclocal22}
\eeq

This is the general expression of the Sagnac time delay in the  {interferometer} frame. We see that the Sagnac Effect depends, in general, both (i) on the position of the interferometer in the rotating frame through the acceleration $\bm{\mathcal A}$, whose expression is related to the laboratory location on the Earth, and (ii) on the interferometer size, since the integrands in (\ref{eq:sagnaclocal22}) are not constant across the interferometer area. For a terrestrial laboratory, the leading effect is due to the diurnal rotation (see e.g. \cite{GINGER11}), and the gravitational corrections are some 9 order of magnitude smaller.  Since we are interested to first order expressions in $|\mb x|$, we can neglect the denominator in the first integrand. Moreover, taking into account the expression (\ref{eq:Alab}) of the laboratory acceleration, the terms $\bm \Omega \left( \bm{\mathcal A} \cdot \mb x \right)$ and $\mb x \left(\bm{\mathcal A} \cdot \bm \Omega \right)$ have order of magnitude $\displaystyle \Omega_{\oplus} \frac{GM}{R_{}}\frac{L}{R_{}}$  where $L$ is the linear size of the interferometer, that is a factor $L/R_{}$ smaller than the leading gravitational contribution (see below). Eventually, since $\displaystyle \bm{\mathcal{A}}= - \Omega^{2}_{\oplus} \, R_{} \, \sin \theta $, where $\theta$ is the laboratory colatitude, it is possible to show \cite{sagnac14} that the kinematics corrections non linear in $\Omega_{\oplus}$, can be safely neglected, since $\displaystyle \bm{\mathcal A} \cdot \mb x \simeq {\Omega_{\oplus}^{2} R_{}L}{}  \simeq 4 \times 10^{-19} \left(\frac{L}{\mathrm{1 \ m}} \right)$,  where $L$ is the linear size of the interferometer.   
Then, on choosing the origin in correspondence of the observer (i.e. $\mb{x}_P=0$) and restoring physical units, we may write the time delay in the form
\beq
\Delta \tau =  -\frac{4}{c^{2}}  \int _{S} \bm \Omega \cdot d \mb S =-\frac{4\bm \Omega \cdot \bm S}{c^{2}}  \ \label{eq:sagnaclocal0}
\eeq
which has the same form of the original Sagnac formula (\ref{eq:sagnac2}), where now $\bm \Omega$ contains both the purely kinematical and the gravitational contributions:  $\bm \Omega=-\bm \Omega_{\oplus}-\bm \Omega'$, so that we may write
\beq
\Delta \tau= \frac{4  \bm \Omega_{\oplus} \cdot \mathbf S}{c^{2}}+ \frac{4 \bm \Omega' \cdot \mathbf S}{c^{2}},\label{eq:ringlocal1}
\eeq
In particular, we see that $\displaystyle \frac{4  \bm \Omega_{\oplus} \cdot \mathbf S}{c^{2}}$ is the purely kinematic Sagnac term, due to the rotation of the Earth, while $ \displaystyle \frac{4 \bm \Omega' \cdot \mathbf S}{c^{2}}$ is the gravitational correction due to the contributions (\ref{eq:OmegaDS})-(\ref{eq:OmegaTh}). 

In order to get a further insight into Eqs. (\ref{eq:OmegaDS})-(\ref{eq:OmegaTh}) it is useful to use an
orthonormal spherical basis $\bm{u}_{r}, \bm{u}_{\vartheta}, \bm{u}_{\varphi}$ in the ECI frame, such
that the $\vartheta=\pi/2$ plane coincides with the equatorial plane. As a consequence, we may write the position
vector of the laboratory  $\displaystyle \bm{R}=R \bm{u}_{r}$ with respect to the center of the Earth, and the kinematic constraint $ \displaystyle \bm V= \bm\Omega_{\oplus} \wedge \bm R$, i.e. $\bm V= \Omega_{\oplus} R \sin \theta  \bm u_{\varphi}$.

Accordingly, we may write the components of $\bm \Omega'$ in physical units as
\begin{eqnarray}
\bm \Omega_{G}&=& -\left( 1+\gamma \right)\frac{GM}{c^{2}R} \sin \vartheta   \Omega_{\oplus} \bm{u}_{\vartheta},
\label{eq:OmegaDS10} \\
\bm \Omega_{B}&=&-\frac{1+\gamma +\alpha_{1}/4}{2} \frac{G}{c^{2}R^{3}}\left[\bm J_{\oplus}-3 \left(\bm J_{\oplus} \cdot \bm{u}_{r} \right) \bm{u}_{r} \right],
\label{eq:OmegaLT10} \\
\bm \Omega_{W} & = & -\frac{\alpha_{1}}{4} \frac{GM}{c^{2}R^{2}} \bm u_{r}   \wedge \bm W, \label{eq:Omegaw10} \\
\bm \Omega_{T}&=&-\frac{1}{2c^{2}} \Omega^{2}_{\oplus} R^{2} \sin^{2} \vartheta \bm{\Omega}_{\oplus},
\label{eq:OmegaTh10}
\end{eqnarray}

If we assume the GR values of the PPN parameters ($\gamma=1, \ \alpha_1=0$) and use for
the Newtonian potential of the Earth its monopole approximation ($U(R)=GM/R$), we may explicitly write the total rotation rate  that enters the Eq. (\ref{eq:ringlocal1}):
\begin{eqnarray}
\bm{\Omega}&=& -\bm{\Omega}_{\oplus} +2\frac{GM}{c^{2}R} \sin \vartheta \Omega_{\oplus} \bm{u}_{\theta}+
\frac{G}{c^{2}R^{3}}\left[\bm J_{\oplus}-3 \left(\bm J_{\oplus} \cdot \bm{u}_{r} \right) \bm{u}_{r} \right] \nonumber \\
& & \label{eq:Omegalocal1}
\end{eqnarray}

If we denote by $\alpha$ the angle between the radial direction $\bm u_{r}$ and the normal vector $\bm u_{n}$, on setting $\bm u_{n}=\cos \alpha \bm u_{r}+ \sin \alpha \bm u_{\theta}$ in (\ref{eq:ringlocal1}), and using (\ref{eq:Omegalocal1}), we may express the proper-time delay in the form
\beq
\Delta \tau=\frac{4S}{c^{2}}\left[  \Omega_{\oplus} \cos \left(\theta+\alpha \right)
-2\frac{GM}{c^{2}R}\Omega_{\oplus}\sin \theta \sin \alpha +\frac{GI_{\oplus}}{c^{2}R^{3}}\Omega_{\oplus}  \left(2 \cos \theta \cos \alpha+\sin\theta \sin \alpha \right)  \right] \label{eq:delayexp1}
\eeq
where we have written $\bm J_{\oplus}=I_{\oplus} \bm \Omega_{\oplus}$, in term of the $I_{\oplus}$,  the moment of inertia of the Earth. Since $I_{\oplus} \simeq M R^{2}$, we see that \textit{on the Earth} the de Sitter and the gravito-magnetic contribution have the same order of magnitude. In particular, the gravitational contributions are approximately $10^{-9}$ smaller than the kinematical leading term.

\section{Sagnac Effect in Alternative Theories of Gravity}\label{sec:altSagnac}

 {In this Section we consider the impact that alternative theories of gravity may have in Sagnac-like experiments performed on the Earth. Indeed,  our results should be considered just as preliminary estimates of the effects that the theories considered may have in such experiments: the actual possibility of measuring these effects depend on both their magnitude and on the experimental accuracy that devices like GINGER will be able to achieve.} Theories of gravity alternative tho GR have been proposed for several motivations: for instance, for solving the problems that arise when Einstein's theory is used to explain the observations on galactic and cosmological scale, or to formulate a quantum theory of gravity (see e.g. \cite{Will:2014bqa},\cite{Berti:2015itd}). Some of these theories, in particular, introduce corrections to  Sagnac Effect (\ref{eq:delayexp1}) in GR. Indeed, we have already obtained  the Sagnac time delay for a  wide class of alternative theories of gravity described by the Parameterized Post-Newtonian (PPN) formalism: the rotation rate  is given by Eqs. (\ref{eq:OmegaDS10})-(\ref{eq:OmegaTh10}), and the $\alpha_{1}$ and $\gamma$ PPN parameters can be constrained in this kind of experiments.

The corrections to the Sagnac time delay for the GINGER experiment have been recently calculated \cite{ninfa} in the framework of Horava-Lifshits gravity; the latter, 
is a four-dimensional theory of gravity which is power-counting renormalizable and, hence, can be considered as a candidate for the ultraviolet completion of GR. In particular, the gravitational contributions in Eq. (\ref{eq:delayexp1}) are modified according to 
\beq 
-2\frac{GM}{c^{2}R}\Omega_{\oplus}\sin \theta \sin \alpha \rightarrow \left(1+\frac{G^{*}}{G} a_1-\frac{a_2}{a_1}\right)\frac{G^{*} M}{c^2 R}\sin{\theta}\sin{\alpha} \label{HL1}
\eeq
and
\beq
\frac{GI_{\oplus}}{c^{2}R^{3}}\Omega_{\oplus}  \left(2 \cos \theta \cos \alpha+\sin\theta \sin \alpha \right) \rightarrow \frac{G^{*}I_{\oplus}}{c^{2}R^{3}}\Omega_{\oplus}  \left(2 \cos \theta \cos \alpha+\sin\theta \sin \alpha \right)
\label{HL2}
\eeq
In the above equations $a_{1},a_{2}$ are coupling constants of the theory and $G^{*}$ is the Newtonian constant in the Horava-Liffhits theory, that could, in principle, differ from the GR one.

The Sagnac Effect in conformal Weyl gravity has been studied in \cite{sultana}; the interest in this theory of gravity is due to the capability of explaining the observation of the rotation curves of the galaxies without requiring dark matter. There are additional gravitational contributions to the time delay that depend on the theory parameter $\xi$;  {for instance, for  beams propagating in equatorial orbits around the Earth, these contributions are} 
\beq
\Delta \tau_{W} \simeq  \frac{4\pi R a  \xi}{c}-\frac{2\pi R^{3} \Omega_{\oplus} \xi}{c^{2}} \label{eq:corweyl}
\eeq
where $a=J/Mc$, is the angular momentum per unit mass of the Earth. Actually, on substituting  the relevant data for  {the Earth, the above expression  (\ref{eq:corweyl}) can be used to show that the corrections  } are some 16 orders of magnitude smaller than the leading terrestrial kinematical effect, so well below the capability of current experiments, included GINGER. This is not surprising, since Weyl gravity is significantly different from GR at large scales, so that its corrections are negligible in the case of terrestrial experiments. 

In \cite{Capozziello:2014mea} the impact of extended theories of gravity on the space experiments GP-B and LARES has been evaluated. In these theories, GR is \textit{extended}  on geometric grounds, in order to obtain further degrees of freedom (related to higher order terms, non minimal couplings and scalar fields in the field equations) that can explain observations at large scales. In particular, the authors consider the weak field limit (in order to describe with sufficient accuracy the weak field of the Earth) of a generic scalar-tensor-higher-order model to set constraints deriving from the available and forthcoming data from GP-B and LARES missions. Both the geodetic and  the Lense-Thirring precession terms are modified, by  somewhat complicated combinations of terms that depend on the effective masses $m_{R}, m_{Y}, m_{\phi}$ of the model. In particular, $\bm \Omega_{G} \rightarrow \bm \Omega_{G}+\bm \Omega_{G}^{EG}$, where
\begin{eqnarray}\label{eq:OmegaGEG}
&&\mathbf{\Omega}_{G}^\text{(EG)}\,=\,-\biggl[g(\xi,\eta)(m_R\tilde{k}_Rr+1)\,F(m_R\tilde{k}_R\mathcal{R})\,e^{-m_R\tilde{k}_Rr}+
\frac{8}{3}(m_Yr+1)\,F(m_Y\mathcal{R})\,e^{-m_Yr}\\\nonumber
&&\qquad\qquad\qquad\qquad\qquad\qquad\qquad\qquad +\frac{1}{3}-g(\xi,\eta)](m_R\tilde{k}_\phi r+1)F(m_R\tilde{k}_\phi\mathcal{R})\,e^{-m_R\tilde{k}_\phi r}\biggr]\frac{{\bf \Omega}_{G }^{}}{3}~.
\end{eqnarray}
and $\bm \Omega_{B} \rightarrow \bm \Omega_{B}+\bm \Omega_{B}^{EG}$, where
\begin{eqnarray}
\label{OmegaLTEG}
&&\mathbf{\Omega}_{B}^\text{(EG)}\,=\,-e^{-m_Y r}(1+m_Y r+{m_Y}^2r^2)\,\mathbf{\Omega}_{B}~,\nonumber
\end{eqnarray}
in the above equations $r$ is  the distance, from the center of the Earth, where experiments are performed: so, in the case of a terrestrial laboratory, $r=R$. 

In the Standard Model Extension  (SME) violations of Lorentz symmetry are allowed for both gravity and electromagnetism: actually, these violations could be signals of new physics effects deriving from a still unknown underlying quantum theory of gravity \cite{Bailey1}.  There are 9 coefficients $\bar{s}^{\mu\nu}$ that  parameterize the  effects of Lorentz violation in the gravitational sector, under the assumption of spontaneous Lorentz-symmetry breaking. In particular, in \cite{Bailey2} (see also \cite{Iorio:2012gr})  it is shown that additional contributions deriving from Lorentz violation are present in the gravitational field of a point-like source of mass $M$

 {\begin{eqnarray}
G_{00} &=& 1-2U(R) 
\left[ 1+\frac 32 \bar{s}^{00}+\frac 12 \bar{s}^{jk}\hat X^j \hat X^k \right],
\label{eq:gmodSME1}\\
G_{0j} &=& -U(R) 
\left[ \bar{s}^{0j}+\bar{s}^{0k}\hat X^k \hat X^j \right].
\label{eq:gmodSME2}
\end{eqnarray}
where\footnote{Indeed, the potentials (\ref{eq:gmodSME1})-(\ref{eq:gmodSME2}) are those of a static mass, further contributions can be obtained if the rotation is taken into account.} $U(R)=GM/R$ and $\hat X= \bm R/|\mb R|$}. We have modifications of the de Sitter contribution, due to (\ref{eq:gmodSME1}), and of the gravito-magnetic contribution, due to (\ref{eq:gmodSME2}).

Even though further details need to be clarified, we see that an accurate measurements of the Sagnac Effect could help to set constraints on SME and, as well, on Horava-Lifhists gravity and extended theory of gravity.

\section{GINGER: a Ring Laser Array for Testing Gravity on the Earth} \label{sec:disconc}

\begin{figure}[here]
\begin{center}
\includegraphics[width=8cm]{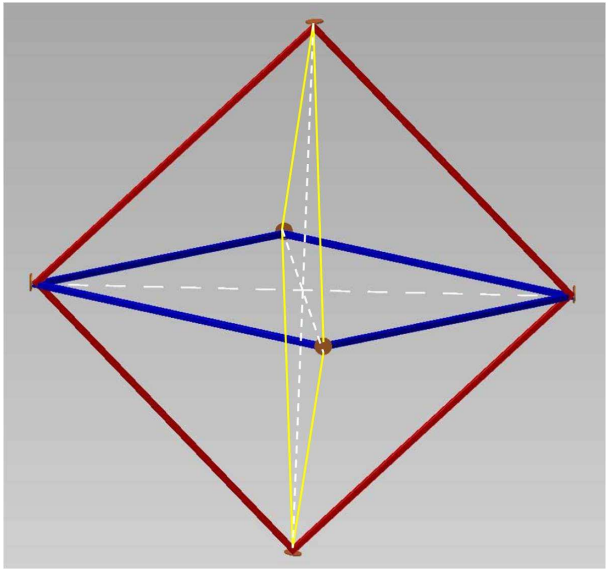}
\caption{ {A possible configuration of GINGER: an octahedron.  Six mirrors give rise to three mutually perpendicular square rings.}}\label{fig:ottaedro}
\end{center}
\end{figure}

As we have seen above, the gravitational contributions to the Sagnac time delay are  much smaller than the leading effect due to the diurnal rotation so, in order to detect these effect,  we need a device with an a accuracy of at least nine order of magnitude 
better than the one required to detect the rotation rate of the Earth. Ring lasers are good candidates for this purpose. 

Let us briefly review how such devices work. A ring laser converts the time
differences (\ref{eq:delayexp1}) into a frequency difference. Indeed, in a ring laser we have continuous and steady light emission;  two standing waves, associated with the two rotation senses, form and
co-exist in the annular cavity of the laser. The asymmetry in the the time of flight difference of the two waves is converted  into different frequencies of the two waves and, eventually, the
frequency difference gives rise to a beat note, which is what can be actually measured. This frequency difference is expressed by the ring laser equation
\begin{equation}
\Delta f=\frac{4}{\lambda P}\ \mathbf{S}_{}\cdot \boldsymbol{\Omega } = \frac{4S}{\lambda P}\ \mathbf{u}_{n}\cdot \boldsymbol{\Omega } 
 ,  \label{eq:ringleq}
\end{equation}
where $P$ is the perimeter and $\lambda $ is the laser wavelength;   $\displaystyle \frac{4S}{\lambda P}$ is the \textit{scale factor} of the device, and is very important in the measurement process.
On substituting from Eq. (\ref{eq:delayexp1}), we have
\beq
\Delta f= \frac{4S}{\lambda P}\left[  \Omega_{\oplus} \cos \left(\theta+\alpha \right)
-2\frac{GM}{c^{2}R}\Omega_{\oplus}\sin \theta \sin \alpha +\frac{GI_{\oplus}}{c^{2}R^{3}}\Omega_{\oplus}  \left(2 \cos \theta \cos \alpha+\sin\theta \sin \alpha \right)  \right]
\label{eq:ringGR}
\eeq
 Actually, in order to detect such tiny effects, the ratio $\frac{4S}{\lambda P}$ must be known and kept at $%
10^{-10}$ accuracy level for the whole measurement period.

What about the available accuracy of ring lasers, today? Commercial ring lasers, that are used for instance in navigation applications, are small devices  with  an accuracy of some  $5\times
10^{-7}$ rad$/$s$/\sqrt{\text{Hz}}$, which is clearly not sufficient for our purposes. 

The starting point for the design and building of GINGER is the Gross Ring (G) in Wettzell \cite{ulli}, which is made of a square ring, 4 m in side, mounted on an
extremely rigid and thermally stable monolithic zerodur slab, located under
an artificial 35 m thick mound. The most recent performance of G, expressed
in terms of measured equivalent angular velocity, has a lower boundary below
$1$ prad/s (picoradian/second) at 1000 s integration time. Even though this accuracy is still not sufficient for the measurement of the gravitational effects,  suitable improvements should help to fill the gap. 

Let us briefly mention the main challenges of this project (we refer to \cite{CR14} for further details and for the road map of the GINGER project).

In order to reach such a demanding accuracy level, there are lots of hypotheses that need to be taken under control. For instance we assumed in Eq. (\ref{eq:ringGR}) that the terrestrial rotation rate is constant, but this is not strictly true,  because of  both the change of the moment of inertia due to the interaction of other celestial bodies and  the actual non rigidity of the Earth. Moreover, the angles appearing in (\ref{eq:ringGR}) are not
stable at the required accuracy of nrad or less, due to the non rigidity
of the laboratory located on the Earth crust and  to thermal instabilities. 

One manifest problem is due to the fact that the GR effects are a very small constant quantity that is always superposed to a huge signal (the kinematical Sagnac term), which makes calibration a difficult task. As a consequence,  an accurate investigation of
the systematics of the laser is needed, and different techniques for
extracting the signal need be considered and evaluated.

Another important issue is that effective rotation along different directions contribute to the beat frequency: consequently, in order to completely measure and distinguish the various terms, it is necessary to have a three-dimensional
device able to measure the three components of the rotation vectors. 
The solution proposed by the GINGER project is given by  a three-dimensional array of square rings (each of which bigger than the
present G ring), mounted on a heterolitic structure, in which the control of the shape of the ring is achieved by  dynamical control of each perimeter.  {A possible configuration for GINGER is shown in Figure (\ref{fig:ottaedro}).
Actually the octahedral structure is the most compact, and, in principle,
easy to control, configuration:  the control is  obtained by means of laser
cavities along the three main diagonals of the octahedron. The side of each
of the three square loops would be not less than $6$ m.} Moreover, the choice of the laboratory location is important, in order to minimize the noise due to  atmospheric phenomena: the proposal is to perform the experiment in a cavern at  the  Gran Sasso underground laboratory
(LNGS), in Italy. 

\section{Conclusions} \label{sec:conc}

In the first part of the paper, we have seen, starting from the very basic principles of relativity, how the Sagnac time delay arises, in arbitrary stationary space-time. In particular, we have shown its universality: it is a geometrical consequence of the space-time structure, and it is not related to the physical nature of the propagating beams, provided that the speed of the particles is a function only of the position along the interferometer path. Then, we have focused on the issue of the analogy with the Aharonov-Bohm: we have written an exact expression of the Sagnac Effect in terms of surface integrals across the interferometer area,  which has  also enabled us to investigate the role of the  position and extensions of the interferometer. We have seen that:  (i) in general, the Sagnac Effect is influenced by both the position of the interferometer in the rotating frame and its extension; (ii)  the analogy with Aharonov-Bohm effect holds true to lowest approximation order only. However, in actual experimental situations, the higher order corrections are negligible and the effect is safely described by the expression (\ref{eq:sagnac2}), both for matter and light beams. In this approximation the analogy with the Aharonov-Bohm effect can be applied, even though the two effects are quite different in general.

Then, we have  described the proper reference frame of a terrestrial laboratory, where measurements are performed by a Sagnac interferometer. In particular, we have obtained the expression of the Sagnac time delay for an arbitrary starting from a PPN metric, describing the gravitational field of the Earth. In particular, we have seen that in GR, for an experiment performed in a terrestrial laboratory, the de Sitter and the gravito-magnetic contribution have the same order of magnitude and are  approximately $10^{-9}$ smaller than the kinematical leading term. 

The main purpose of the GINGER project is the detection of these very small gravitational effects, for the first time, in a terrestrial laboratory: the device for reaching this goal is a three-dimensional array of advanced ring lasers.  Besides measuring the predictions of General Relativity, it should be possible to perform high precision tests of metric theories of gravity in the framework of the PPN formalism, also taking into account the possible improvements and upgrading of the apparatus. We have seen that also theories of gravity that falls beyond the PPN formalism have an impact on Sagnac Effect, so GINGER may help to set constraints, for instance, on Horava-Lifshits gravity, extended theories of gravity, Standard Model Extension. 

The  measurements performed by a ring laser require a multidisciplinary approach, involving geodesy and geophysics. In particular, GINGER together with other similar devices in the world, could provide a very accurate measurement of the terrestrial location. Moreover, since such a device is a very accurate inertial sensor, it could be useful in geophysics, in particular for rotational seismology. 

In summary, the task of projecting and realizing a similar project is quite demanding, however this effort will be compensated by the importance and uniqueness of the experiment. In the end, it is worthwhile emphasizing that the fundamental idea was conceived by Sagnac, about 100 years ago, and it exploits light as a probe of gravity: one more reason to stress the importance of  light and optical technologies in current research, in this  International Year of Light.










\bibliographystyle{mdpi}
\makeatletter
\renewcommand\@biblabel[1]{#1. }
\makeatother



%


%

\end{document}